\documentclass[12pt,preprint]{aastex}
\def\lsim{\mathrel{\rlap{\lower4pt\hbox{\hskip1pt$\sim$}}
    \raise1pt\hbox{$<$}}}                
\def\gsim{\mathrel{\rlap{\lower4pt\hbox{\hskip1pt$\sim$}}
    \raise1pt\hbox{$>$}}}                

\shorttitle{Statistical isotropy of CMB}
\shortauthors{Hajian and Souradeep}
\slugcomment{IUCAA-37/2003}

\begin{document}

\bibliographystyle{apj} 

\title{Measuring Statistical isotropy of the CMB anisotropy}
\author{Amir Hajian and Tarun Souradeep } \affil{Inter-University
Centre for Astronomy and Astrophysics,\\ Post Bag 4, Ganeshkhind, Pune
411007, India}

\begin{abstract}

  
  The statistical expectation values of the temperature fluctuations
  of cosmic microwave background (CMB) are assumed to be preserved
  under rotations of the sky. This assumption of {\em statistical
  isotropy} (SI) of the CMB anisotropy should be observationally
  verified since detection of violation of SI could have profound
  implications for cosmology.  We propose a set of measures,
  $\kappa_\ell$ ($\ell=1,2,3, \ldots$) for detecting violation of
  statistical isotropy in an observed CMB anisotropy sky map indicated
  by non zero $\kappa_\ell$.  We define an estimator for the
  $\kappa_\ell$ spectrum and analytically compute its cosmic bias and
  cosmic variance.  The results match those obtained by measuring
  $\kappa_\ell$ using simulated sky maps. Non-zero (bias corrected)
  $\kappa_\ell$ larger than the SI cosmic variance will imply
  violation of SI. The SI measure proposed in this paper is an
  appropriate statistics to investigate preliminary indication of SI
  violation in the recently released WMAP data.

\end{abstract}


The Cosmic Microwave Background (CMB) anisotropy is a very powerful
observational probe of cosmology. In standard cosmology, CMB
anisotropy is expected to be statistically isotropic, i.e.,
statistical expectation values of the temperature fluctuations $\Delta
T(\hat q)$ are preserved under rotations of the sky. In particular,
the angular correlation function $C(\hat{q},\,
\hat{q}^\prime)\equiv\langle\Delta T(\hat q)\Delta T(\hat
q^\prime)\rangle$ is rotationally invariant for Gaussian fields. In
spherical harmonic space, where $\Delta T(\hat q)= \sum_{lm}a_{lm}
Y_{lm}(\hat q)$ this translates to a diagonal $\langle a_{lm}
a^*_{l^\prime m^\prime}\rangle=C_{l}
\delta_{ll^\prime}\delta_{mm^\prime}$ where $C_l$, the widely used
angular power spectrum of CMB anisotropy, is a complete description of
(Gaussian) CMB anisotropy. Hence, it is important to be able to
determine if the observed CMB sky is a realization of a statistically
isotropic process, or not~\footnote{ Statistical isotropy of CMB
anisotropy and its measurement has been discussed in literature
earlier~\cite{fer_mag97,bun_scot00}.}.

We propose a set of measures, $\kappa_\ell$ ($\ell=1,2,3, \ldots$)
which for non-zero values indicate and quantify statistical isotropy
violation in a CMB map.  A null detection of $\kappa_\ell$ will be a
direct confirmation of the {\em assumed} statistical isotropy of the
CMB sky. It will also justify model comparison based on the angular
power spectrum $C_l$ alone~\cite{bps}.  The detection of statistical
isotropy (SI) violation can have exciting and far-reaching implication
for cosmology.  In particular, SI violation in the CMB anisotropy is
the most generic signature of non-trivial geometrical and topological
structure of space on ultra-large-scales. Non-trivial cosmic topology
is a theoretically well motivated possibility that is being
observationally probed on the largest scales only
recently~\cite{ell71,lac_lum95,stark98, lev02}.

For statistically isotropic CMB sky, the correlation function

\begin{equation}
C(\hat{n}_1,\hat{n}_2)\equiv C(\hat{n}_1\cdot\hat{n}_2) = \frac{1}{8\pi^2}\int d{\mathcal R} 
C({\mathcal R}\hat{n}_1,\, {\mathcal
   R}\hat{n}_2),
\label{avg_cth}
\end{equation}
where ${\mathcal R}\hat{n}$ denotes the direction obtained under the
action of a rotation ${\mathcal R}$ on $\hat{n}$, and $d{\mathcal R}$
is a volume element of the three-dimensional rotation group.  The
invariance of the underlying statistics under rotation allows the
estimation of $C(\hat{n}_1\cdot\hat{n}_2)$ using the average of the
temperature product $\widetilde{\Delta T}(\hat n) \widetilde{\Delta
T}(\hat n')$ between all pairs of pixels with the angular separation
$\theta$.  In particular, for CMB temperature map $\widetilde{\Delta
T}(\hat q_i)$ defined on a discrete set of points on celestial sphere
(pixels) $\hat q_i$ ($i=1,\ldots,N_p$)

\begin{equation}
\tilde C(\theta)\,=\,
\sum_{i,j=1}^{N_p}  \widetilde{\Delta T}(\hat q_i) \widetilde{\Delta T}(\hat q_j) \delta( \cos\theta - \hat q_i  \cdot \hat q_j)\,,  
\label{bin_cth}
\end{equation}
is an estimator of the correlation function $C(\theta)$ of an
underlying SI statistics~\footnote{ This simplified description does
not include optimal weights to account for observational issues, such
as instrument noise and non uniform coverage. However, this is well
studied in literature and we evade these to keep our presentation
clear.}.

In the absence of statistical isotropy, $C(\hat{q},\hat{q}')$ is
estimated by a single product $\widetilde{\Delta T}(\hat
q)\widetilde{\Delta T}(\hat q')$ and hence is poorly determined from a
single realization. Although it is not possible to estimate each
element of the full correlation function $C(\hat{q},\hat{q}')$, some
measures of statistical anisotropy of the CMB map can be estimated
through suitably weighted angular averages of $\widetilde{\Delta
T}(\hat q)\widetilde{\Delta T}(\hat q')$. The angular averaging
procedure should be such that the measure involves averaging over
sufficient number of independent `measurements', but should ensure
that the averaging does not erase all the signature of statistical
anisotropy (as would happen in eq.~(\ref{avg_cth}) or
eq.~(\ref{bin_cth})). Another important desirable property is that the
measures be independent of the overall orientation of the sky. Based
on these considerations, we propose a set of measures $\kappa_\ell$ of
statistical isotropy violation given by
\begin{equation}\label{kl}
 \kappa_\ell= \int\!\!  d\Omega\!\!\int\!\!  d\Omega^\prime 
 \left[\frac{(2\ell+1)}{8\pi^2}\!\!\int d{\mathcal R} \chi_\ell({\mathcal R})
C({\mathcal R}\hat{q},{\mathcal R}\hat{q}^\prime)\right]^2\!\!\!\!,
\end{equation}
where $ C({\mathcal R}\hat{q},\, {\mathcal R}\hat{q}^\prime)$ is the
two point correlation between ${\mathcal R}\hat{q}\,$ and $ {\mathcal
R}\hat{q}^\prime$ obtained by rotating $\hat{q}$ and $\hat{q}^\prime$
by an element ${\mathcal R}$ of the rotation group.  The measures
$\kappa_\ell$ involve angular average of the correlation weighed by
the characteristic function of the rotation group $\chi_\ell({\mathcal
R}) =\sum_{M} D_{MM}^{\ell}({\mathcal R})$ where $
D_{MM^\prime}^{\ell}$ are the Wigner D-functions~\cite{Var}.  When
${\mathcal R}$ is expressed as rotation by an angle $\omega$ ( where
$0\leq \omega \leq \pi$) about an axis ${\hat r}(\Theta, \Phi)$, the
characteristic function $ \chi_\ell({\mathcal
R})\,\equiv\,\chi_\ell(\omega)
=\sin{[(2\ell+1)\omega/2]}/{\sin{[\omega/2]}}$ is completely
determined by $\omega$ and the volume element of the three-dimensional
rotation group is given by $ d{\mathcal R}\,=\, 4 \, \sin^2{\omega
/{2}}\, d\omega \, \sin{\Theta}\, d\Theta \, d\Phi\,$. Using the
identity $\int d{\mathcal R^\prime} \chi_\ell({\mathcal
R^\prime})\chi_\ell({\mathcal R R^\prime}) =\chi_\ell({\mathcal R})$,
expression (\ref{kl}) can be simplified to
\begin{equation}\label{klsimp}
 \kappa_\ell= \frac{(2\ell+1)}{8\pi^2} 
\int\!\!  d\Omega\!\!\int\!\!  d\Omega^\prime C(\hat{q},\hat{q}^\prime)
 \!\!\int d{\mathcal R} \chi_\ell({\mathcal R})
C({\mathcal R}\hat{q},{\mathcal R}\hat{q}^\prime)\!\!\!\!,
\end{equation}
containing only one integral over the rotation group.  For a
statistically isotropic model $ C({\mathcal R}\hat{q}_1,\, {\mathcal
R}\hat{q}_2)\,\equiv\,C(\hat{q}_1,\, \hat{q}_2)$ is invariant under
rotation, and eq.~(\ref{klsimp}) gives $\kappa_\ell \, = \, \kappa^0
\delta_{\ell 0},$ due to the orthonormality of
$\chi_\ell(\omega)$. Hence, $\kappa_\ell$ defined in eq.~(\ref{kl}) is
a measure of statistical isotropy.

The measure $\kappa_\ell$ has clear interpretation in harmonic space.
The two point correlation $C(\hat{q},\, \hat{q}^\prime)$ can be
expanded in terms of the orthonormal set of bipolar spherical
harmonics as
\begin{equation}\label{bipolar}
 C(\hat{q},\, \hat{q}^\prime)\, =\, \sum_{ll^\prime\ell M} A_{ll^\prime}^{\ell
  M}
 \{Y_{l}(\hat{q}) \otimes Y_{l^\prime}(\hat{q}^\prime)\}_{\ell M}\,,
\end{equation}
where $A_{ll^\prime}^{\ell M}$ are the coefficients of the expansion.
These coefficients are related to `angular momentum' sum over the
covariances $\langle a_{lm}a^*_{l^\prime m^\prime}\rangle$ as 
\begin{equation}\label{alml1l2}
 A_{ll^\prime}^{\ell M} = 
\sum_{m m^\prime} \langle a_{lm}a^*_{l^\prime m^\prime}\rangle
\, \, (-1)^{m^\prime} \mathfrak{ C}^{\ell M}_{lml^\prime -m^\prime}\,,
\end{equation}
where $\mathfrak{C}^{\ell M}_{lml^\prime m^\prime}$ are Clebsch-Gordan
coefficients.  The bipolar functions transform just like ordinary
spherical harmonic function $Y_{\ell M}$ under rotation~\cite{Var}.
Substituting the expansion eq.~(\ref{bipolar}) into eq.~(\ref{kl}) we
can show that
\begin{equation}\label{klharm}
\kappa_\ell =\sum_{ll^\prime M} |A_{ll^\prime}^{\ell M}|^2 \geq 0\,,
\end{equation}
is positive semidefinite and can be expressed in the form
\begin{equation}
 \kappa_\ell = \frac{2\ell+1}{8\pi^2}\!\!
\int\!\! d{\mathcal R}\chi_\ell({\mathcal R})
\sum_{lml^\prime m^\prime}\langle a_{lm}a^*_{l^\prime m^\prime}
\rangle\langle a_{lm}a^*_{l^\prime m^\prime}
\rangle^{\mathcal R}\!\!,
\end{equation}
where $\langle\ldots\rangle^{\mathcal R}$ is computed in a frame
rotated by ${\mathcal R}$.  When SI holds $\langle a_{lm}a^*_{l^\prime
m^\prime}\rangle=C_{l}\delta_{ll^\prime}\delta_{mm^\prime}$, implying
$A_{ll^\prime}^{\ell M}=(-1)^l C_{l} (2l+1)^{1/2} \,
\delta_{ll^\prime}\, \delta_{\ell 0}\, \delta_{M0}$. The coefficients
$A_{ll}^{00}$ represent the statistically isotropic part of a general
correlation function. They also represent the statistically isotropic
part of any arbitrary correlation function.  The coefficients
$A_{l_1l_2}^{\ell M}$ are inverse-transform of the two point
correlation
\begin{equation}\label{Alml1l2}
 A_{l_1l_2}^{\ell M} \,=\,\int d\Omega\int d\Omega^\prime \,
 C(\hat{n},\, \hat{n}^\prime)\, \{Y_{l_1}(\hat{n}) \otimes 
Y_{l_2}(\hat{n}^\prime)\}_{\ell M}^*. 
\end{equation}
The symmetry  $C(\hat{n},\hat{n}^\prime) = C(\hat{n}^\prime,\hat{n})$
implies 
\begin{equation} \label{sym}
 A_{l_2l_1}^{\ell M}\,=\,(-1)^{(l_1+l_2-\ell)}A_{l_1l_2}^{\ell M}, \quad
 A_{ll}^{\ell M} \, =\, A_{ll}^{\ell M} \,\,\delta_{\ell,2k}, \quad  
k=0,\,1,\,2,\,\cdots.
\end{equation}

Recently, the Wilkinson Microwave Anisotropy Probe (WMAP) has provided
high resolution (almost) full sky maps of CMB anisotropy~\cite{wmap}
from which $\kappa_\ell$ can be measured.  Given a single independent
CMB map, $\widetilde{\Delta T}(\hat{q})$ we need to look for violation
of statistical isotropy.  Formally, the estimation procedure involves
averaging the product of temperature at pairs of pixels obtained by
rotating a given pair of pixels by an angle $\omega$ around a
sufficiently large sample of rotation axes.  The integral in the
braces in eq.~(\ref{kl}) is estimated by summing up the terms for
different values of $\omega$ weighed by the characteristic function.
We can define an estimator for $\kappa_\ell$ as

\begin{eqnarray}\label{klest}
\tilde\kappa_\ell &&= \tilde\kappa_\ell^B - {\mathfrak B}_\ell,\nonumber \\
\tilde\kappa_\ell^B &&= \frac{(2\ell+1)}{8\pi^2} 
\sum_{i,j=1}^{N_p} \widetilde{\Delta T}(\hat{q}_i)\,\widetilde{\Delta T}(\hat{q}_j)
\sum_{m=1}^{N_w} \chi_\ell(w_m) \sum_{n=1}^{N_r}
\widetilde{\Delta T}({\mathcal R}_{mn}\hat{q}_i)\widetilde{\Delta T}({\mathcal R}_{mn}\hat{q}_j)\,,
\end{eqnarray}
where as described below ${\mathfrak B}_\ell\equiv
\langle\tilde\kappa_\ell^B\rangle$ accounts for the `cosmic bias' for
the biased estimator $\tilde\kappa_\ell^B$.  As with the sky, the
rotation group is also discretized as ${\mathcal R}_{mn}$ where $m
=1,\ldots,N_w$ is an index of equally spaced intervals in rotation
angle $w$ and $n=1,\ldots,N_r$ indexes a set of equally spaced
directions in the sky. While we have also implemented this real space
computation, practically, we find it faster to estimate $\kappa_\ell$
in the harmonic space by taking advantage of fast methods of spherical
harmonic transform of the map. In harmonic space, we first define an
unbiased estimator for the bipolar harmonic coefficients based on
eq.~(\ref{alml1l2}) and then estimate $\kappa_\ell$ using
eq.~(\ref{klharm})

\begin{equation}\label{klALMesthar}
\tilde A_{ll^\prime}^{\ell M} = \sum_{m m^\prime} a_{lm}a_{l^\prime
m^\prime} \, \, \mathfrak{ C}^{\ell M}_{lml^\prime m^\prime}\,\quad
\tilde\kappa_\ell =\sum_{ll^\prime M} \left|\tilde A_{ll^\prime}^{\ell
M}\right|^2 - {\mathfrak B}_\ell\,.
\end{equation}

Assuming Gaussian statistics of the temperature fluctuations, the
cosmic bias is given by~\cite{us_prd}

\begin{eqnarray}\label{klbias}
\langle\tilde\kappa_\ell^B\rangle-\kappa_\ell = \sum_{l_1,l_2}
\sum_{m_1,m_1^\prime} \sum_{m_2,m_2^\prime}&&\,\left[\langle
a^*_{l_1m_1}a_{l_1 m_1^\prime}\rangle\langle a^*_{l_2m_2}a_{l_2
m_2^\prime}\rangle + 
\langle a^*_{l_1m_1}a_{l_2 m_2^\prime}\rangle\langle a^*_{l_2m_2}a_{l_1
m_1^\prime}\rangle \right] \nonumber \\
&&{}\times \sum_M \mathfrak{ C}^{\ell
M}_{l_1m_1l_2m_2}\mathfrak{ C}^{\ell M}_{l_1m_1^\prime l_2m_2^\prime}\,.
\end{eqnarray}

Given a single CMB sky-map, the individual elements of the $\langle
a_{lm}a^*_{l^\prime m^\prime}\rangle$ covariance are poorly
determined.  So we can correct for the bias ${\mathfrak B}_\ell$ that
arises from the SI part of correlation function where

\begin{equation}\label{klisobias}
{\mathfrak B}_\ell \equiv\langle\tilde\kappa_\ell^B\rangle_{_{\rm SI}} =
 (2\ell+1)\,\sum_{l_1} \sum_{l_2=|\ell-l_1|}^{\ell+l_1} C_{l_1}
 C_{l_2} \left[1 + (-1)^{\ell}\, \delta_{l_1 l_2}\right]\,.
\end{equation}
Hence, for SI correlation, the estimator $\tilde \kappa_\ell$ is
unbiased, i.e., $\langle \tilde \kappa_\ell \rangle=0$.

Assuming Gaussian CMB anisotropy, the cosmic variance of the
estimators $\tilde A_{ll^\prime}^{\ell M}$ and $\tilde \kappa_\ell$
can be obtained analytically for full sky maps. The cosmic variance of
the bipolar coefficients

\begin{eqnarray}
\sigma^2(\tilde A_{l_1l_2}^{\ell M}) = \sum_{m_1,m_1^\prime}
\sum_{m_2,m_2^\prime}\,&&\left[\langle a_{l_1m_1}a_{l_1
m_1^\prime}\rangle\langle a_{l_2m_2}a_{l_2 m_2^\prime}\rangle +
\langle a_{l_1m_1}a_{l_2 m_2^\prime}\rangle\langle a_{l_2m_2}a_{l_1
m_1^\prime}\rangle \right] \nonumber \\
&&{}\times \mathfrak{ C}^{\ell
M}_{l_1m_1l_2m_2}\mathfrak{ C}^{\ell M}_{l_1m_1^\prime l_2m_2^\prime},
\end{eqnarray}
which, for SI correlation, further simplifies to

\begin{equation}
\sigma^2_{_{\rm SI}}(\tilde A_{l_1l_2}^{\ell M}) = C_{l_1}\, C_{l_2}\,
\left[ 1 + (-1)^\ell \, \delta_{l_1l_2}\right]\, \sum_{m_1,m_2}
(-1)^{m_1+m_2} \mathfrak{ C}^{\ell M}_{l_1m_1l_2m_2}\mathfrak{
C}^{\ell M}_{l_1m_1^\prime l_2m_2^\prime}.
\end{equation}  
Note that for $l_1=l_2$ the cosmic variance is zero for odd $\ell$ due
to eq.~(\ref{sym}) arising from symmetry of $C(\hat q,\hat q^\prime)$.

\begin{figure}[h]
\includegraphics[scale=0.6, angle=-90]{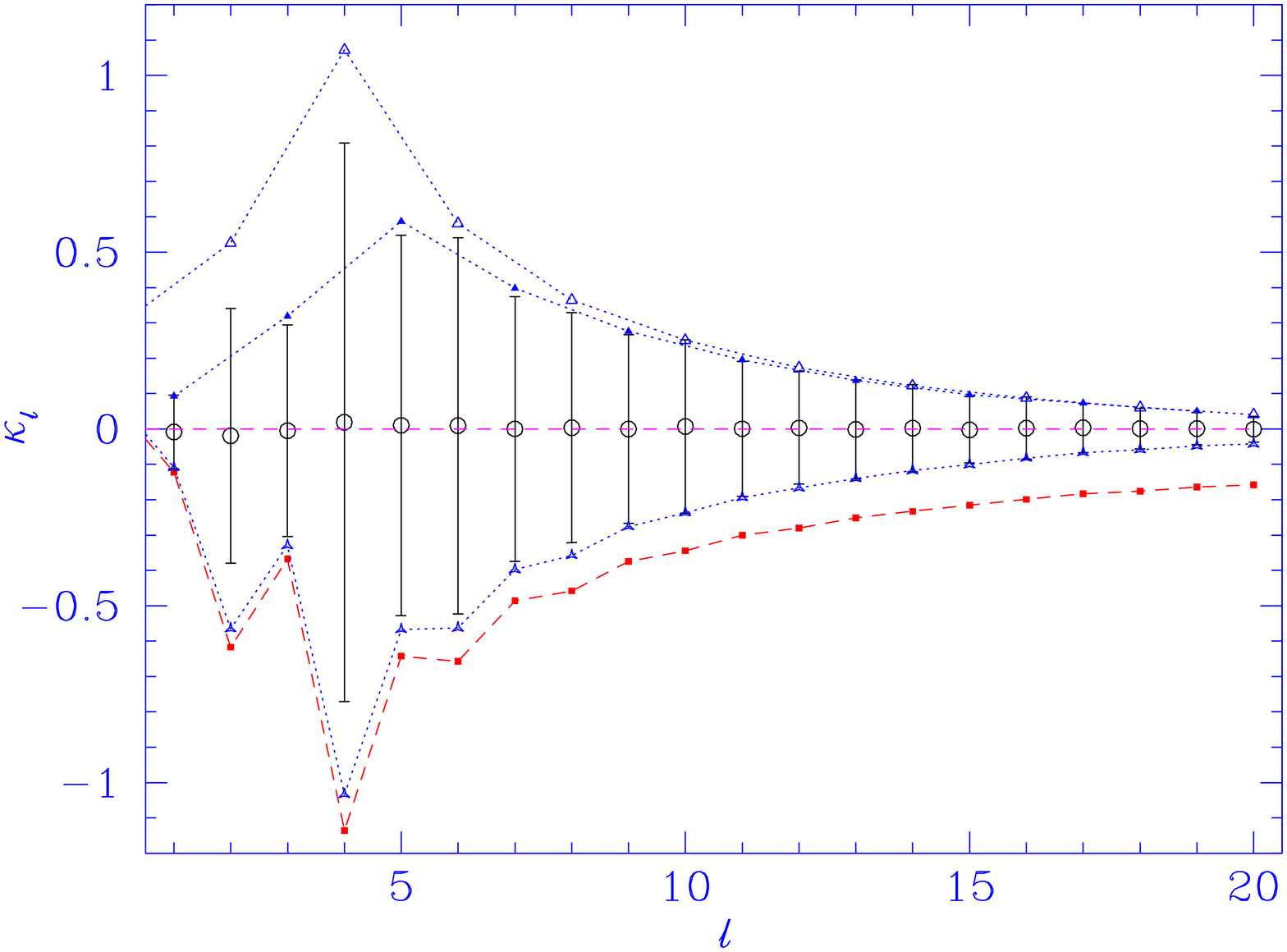}
\caption{ The figure shows the bias corrected `measurement' of
 $\kappa_\ell$ of a SI CMB sky with a flat band power spectrum
 smoothed by a Gaussian beam ($l(l+1) C_l = \exp(-l^2/18^2)$. The
 cosmic error, $\sigma({\kappa_\ell})$, obtained using $50$
 independent realizations of CMB (full) sky map match the analytic
 results shown by lower dotted curve with stars. The upper dotted
 curves separately outline the cosmic error envelope for odd
 multipoles (filled triangles) and for even multipoles (empty
 triangles) to explicitly highlight their convergence. Violation of SI
 will be indicated by non-zero $\kappa_\ell$ measured in an observed
 CMB map in excess of $\sigma({\kappa_\ell})$ given by the $C_l$ of
 the map. The lower dashed curve (filled squares) shows the cosmic
 error for ideal unit flat band power spectrum ($l(l+1) C_l =1$) with
 no beam smoothing. The curve falls off roughly at $1/\ell$ at large
 $\ell$.}
  \label{klmeas}
\end{figure}

A similar but more tedious computation of $105$ terms of the $8$-point
correlation function yields an analytic expression for the cosmic
variance of $\tilde\kappa_\ell$~\cite{us_prd}. For SI correlation, the
cosmic variance for $\ell>0$ is given by

\begin{eqnarray}\label{klcv}
&&\sigma^2_{_{\rm SI}}(\tilde\kappa_\ell) 
=\sum_{l : 2l \ge \ell}\!\! 4\, C_{l}^4 \left[ 2 \frac{(2\ell+1)^2}{2l+1}+ 
(-1)^{\ell} (2\ell+1)+ (1+2(-1)^{\ell}) F_{ll}^\ell\right]
\nonumber \\
&&{}+\sum_{l_1} \sum_{l_2=|\ell-l_1|}^{\ell+l_1} \!\!\! 4
\,C_{l_1}^2\,C_{l_2}^2 \left[ (2\ell+1) + F_{l_1l_2}^\ell \right] + \,8\sum_{l_1}\,\frac{(2\ell+1)^2}{2l_1+1} \,C_{l_1}^2
\left[\sum_{l_2=|\ell-l_1|}^{\ell+l_1} C_{l_2}\right]^2\nonumber \\
&&{} +
16\,(-1)^{\ell}\,\sum_{l_1 : 2l_1 \ge \ell}
\,\frac{(2\ell+1)^2}{2l_1+1}\, \sum_{l_2=|\ell-l_1|}^{\ell+l_1}
 C_{l_1}^3 C_{l_2},
\\{\rm where}   && F_{l_1l_3}^\ell = \!\!\!\!
\sum_{m_1m_2=-l_1}^{l_1}\,\sum_{m_3m_4=-l_3}^{l_3}
\sum_{M,M'=-\ell}^\ell C^{\ell M}_{l_{1}-m_{1}l_{3}-m_{3}}C^{\ell
M}_{l_{1}m_{2}l_{3}m_{4}} C^{\ell M'}_{l_{3}m_{4}l_{1}m_{ 1}}C^{\ell
M'}_{l_{3}-m_{3}l_{1}-m_{2}}.
\end{eqnarray}
Numerically, it is advantageous to rewrite $F_{ll^\prime}^\ell$ in a
series involving $9$-j symbols. The expressions for variance and bias
are valid for full sky CMB maps. For observed maps one has to contend
to incomplete or non uniform sky coverage. In such cases one would
estimate the cosmic bias and variance from averaging over many
independent realizations simulated CMB sky from the same underlying
correlation function. Fig.~\ref{klmeas} shows the measurement of
$\kappa_\ell$ in a SI model with flat band power spectrum. The bias
and variance is estimated from making measurements on $50$ independent
random full-sky maps using the HEALPix~\footnote{Publicly available at
{http://www.eso.org/science/healpix/}}. The cosmic bias and
variance obtained from these realizations match the analytical
results. Just as in the case of cosmic bias, the cosmic variance of
$\kappa_\ell$ at odd multipoles is smaller. The figure clearly shows
that the envelope of cosmic variance for odd and even mulitpole
converge at large $\ell$.  For a constant $l(l+1) C_l$ angular power
spectrum the $\sigma_{_{\rm SI}}(\tilde\kappa_\ell)$ falls off
approximately as $1/\ell$ at large $\ell$. (The absence of dipole and
monopole in the maps affects $\kappa_\ell$ for $\ell < 4$ leading to
the apparent rise in cosmic variance at $\ell < 4$ seen in
Fig~\ref{klmeas}.)

The bias and cosmic variance depends on the total SI angular power
spectrum of the signal and noise $C_l = C_l^S + C_l^N$. Hence, where
possible, prior knowledge of the expected $\kappa_\ell$ signal should
be used to construct multipole space windows to weigh down the
contribution from the region of multipole space where SI violation is
not expected, e.g., the generic breakdown of statistical isotropy due
to cosmic topology. The underlying correlation patterns in the CMB
anisotropy in a multiply connected universe is related to the the
symmetry of Dirichlet domain ~\cite{wol94,vin93}. In a companion
paper, we study the $\kappa_\ell$ signal expected in flat, toroidal
models of the universe and connect the spectrum to the principle
directions in the Dirichlet domain~\cite{us_prl}. SI violation arising
from cosmic topology is usually limited to low multipoles. A wise
detection strategy would be to smooth CMB maps to the low angular
resolution.  When searching for specific form of SI violation, linear
combinations of $\kappa_\ell$ can be used to optimize signal to noise.
Before ascribing the detected breakdown of statistical anisotropy to
cosmological or astrophysical effects, one must carefully account for
and model into the SI simulations other mundane sources of SI
violation in real data, such as, incomplete and non-uniform sky
coverage, beam anisotropy, foreground residuals and statistically
anisotropic noise.

In summary, the $\kappa_\ell$ statistics quantifies breakdown of SI
into a set of numbers that can be measured from the single CMB sky
available.  The $\kappa_\ell$ spectrum can be measured very fast even
for high resolution CMB maps. The statistics has very clear
interpretation as quadratic combinations of off-diagonal correlations
between $a_{lm}$ coefficients. Signal SI violation is related to
underlying correlation patterns. The angular scale on which the
off-diagonal correlations (patterns) occur is reflected in the
$\kappa_\ell$ spectrum.  As a tool for detecting cosmic topology (more
generally, cosmic structures on ultra-large scales), the $\kappa_\ell$
spectrum has the advantage of being independent of overall orientation
of the correlation pattern. This is particularly suited for search for
cosmic topology since the signal is independent of the orientation of
the DD. (However, orientation information is available in the $A^{\ell
M}_{l_1l_2}$.) The recent all sky CMB map from WMAP is an ideal data
set where one can measure statistical isotropy.  Interestingly, there
are hints of SI violation in the low multipole of
WMAP~\cite{maxwmap,angelwmap,erik03}.  Hence is of great interest to
make a careful estimation of SI violation in the WMAP data via
$\kappa_\ell$ spectrum. This work is in progress and results will be
reported elsewhere~\cite{us_wmap}. This approach complement direct
search for signature of cosmic topology~\cite{circles}.

\acknowledgments 

TS acknowledges enlightening discussions with Larry Weaver, Kansas
State University, at the start of this work. TS also benefited from
discussions with J. R. Bond and D. Pogosyan on cosmic topology and related
issues.



\clearpage

\end{document}